\documentclass[traditabstract]{aa} 
\usepackage{graphicx}
\usepackage{txfonts}
\usepackage[]{natbib}
\bibpunct{(}{)}{;}{a}{}{,}
\DeclareMathSymbol{\lesssim}{\mathrel}{AMSa}{"2E}
\usepackage{color}

\begin{document}

\bibliographystyle{aa}
 \title{The puzzling case of the accreting millisecond X--ray pulsar IGR J00291+5934: flaring optical emission during quiescence\thanks{Based on observations made with the Gran Telescopio Canarias (GTC), installed in the Spanish Observatorio del Roque de los Muchachos of the Instituto de Astrofísica de Canarias, in the island of La Palma.}}

   \author{M. C. Baglio
          \inst{1, 2},          
          \         
          S. Campana
          \inst{2}, 
          \
          P. D'Avanzo
          \inst{2}, 
          \
          A. Papitto
          \inst{3},
          \
          L. Burderi
          \inst{4},
          \
          T. Di Salvo
          \inst{5},
          \
          T. Mu\~noz Darias
          \inst{6,7},
          \
          N. Rea
          \inst{8,9},
          \
          D. F. Torres
          \inst{8, 10}
          }
                  
  \institute{Universit\`{a} dell'Insubria, Dipartimento di Scienza e Alta Tecnologia, Via Valleggio 11, I-22100 Como, Italy                      \\
              \email{cristina.baglio@brera.inaf.it}
         \
             \and 
             INAF, Osservatorio Astronomico di Brera, Via E. Bianchi 46, I-23807 Merate, Italy
             \and
             INAF, Osservatorio Astronomico di Roma, Via di Frascati 33, I-00044 Monte Porzio Catone, Italy
             \and
             Dipartimento di Fisica, Universit\`{a} degli Studi di Cagliari, SP Monserrato-Sestu km 0.7, 09042 Monserrato, Italy 
             \and
             Universit\`{a} degli Studi di Palermo, Dipartimento di Fisica e Chimica, via Archirafi 36, I-90123 Palermo, Italy
             \and
             Instituto de Astrof\'{i}sica de Canarias, E-38205 La Laguna, Tenerife, Spain
             \and
             Departamento de Astrof\'{i}sica, Univ. de La Laguna, E-38206 La Laguna, Tenerife, Spain
             \and
             Institute of Space Sciences (CSIC-IEEC), Campus UAB, Carrer Can Magrans s/n, E-08193 Barcelona, Spain
             \and
             Anton Pannekoek Institute for Astronomy, University of Amsterdam, Postbus 94249, NL-1090 GE Amsterdam, The Netherlands 
             \and
             Instituci\'{o} Catalana de Recerca i Estudis Avan\c{c}ats (ICREA) Barcelona, Spain
             }         

   \date{ }

   \abstract{We present an optical ($gri$) study during quiescence of the accreting millisecond X-ray pulsar IGR J00291+5934 performed with the 10.4m Gran Telescopio Canarias (GTC) in August 2014. Although the source was in quiescence at the time of our observations, it showed a strong optical flaring activity, more pronounced at higher frequencies (i.e. the $g$ band). After subtracting
the flares, we tentatively recovered a sinusoidal modulation at the system orbital period in all bands, even when a significant phase shift with respect to an irradiated star, typical of accreting millisecond X-ray pulsars, was detected. We conclude that the observed flaring could be a manifestation of the presence of an accretion disc in the system. The observed light curve variability could be explained by the presence of a superhump, which might be another proof of the formation of an accretion disc. In particular, the disc at the time of our observations was probably preparing the new outburst of the source, which occurred a few months later, in 2015. }

\authorrunning{M. C. Baglio et al.} 
\titlerunning{Optical photometry of IGR J00291+5934 during quiescence }
\maketitle

\section{Introduction}
Accreting millisecond X-ray pulsars (AMXPs) are a subclass of transient low-mass X-ray binaries (LMXBs) hosting a weakly magnetised
fast-spinning (a few ms) neutron star (NS) that accretes matter from a low-mass companion ($ M\lesssim 1M_{\odot} $).
Typically, their spin periods span between 1.7 and 5.4 ms, whereas the orbital periods range between 40 min and 19 hr 
(see \citealt{PatrunoReview} for a review). 

For many years, AMXPs were thought to be the progenitors of millisecond radio pulsars. According to this recycling scenario \citep{Alpar82}, millisecond radio pulsars arise from the spin-up of the neutron stars hosted in LMXBs as a result of the transfer of angular 
momentum (and mass) from their companion stars. After a mass accretion phase, during which the system appears as a bright 
(possibly transient) X-ray source, the mass transfer rate declines and allows for the re-activation of a rotation-powered pulsar that emits from the radio to the gamma-ray band. 
The first AMXP that confirmed this scenario was SAX J1808.4--3658, in which a coherent pulsed signal at 2.5 ms in the X-rays was discovered \citep{Wijnands1998}. Later on, systems alternating within a short time from the AMXP state to the millisecond radio pulsar state were discovered 
(named transitional pulsars), which fully validated the picture (\citealt{Archibald2009}; \citealt{Papitto2013};  \citealt{DeMartino2013}; \citealt{Bassa2014}; \citealt{Roy2015}). 

Optical observations of AMXPs during quiescence are the only way to study the companion star, since during outbursts its emission is otherwise overwhelmed by that of the accretion disc. The optical light curve resulting from these
observations is usually modulated at the orbital period of the source. When the light curve is dominated by the elongated, Roche-lobe deformed
geometry of the companion star,
a double-humped light curve is observed with two equal maxima at the descending and ascending nodes of the companion star (orbital phases 0.25 and 0.75, respectively) and two unequal minima at superior and inferior conjunction (orbital phases 0 ad 0.5, respectively). In systems with small orbital separations, however, the light curve can possess  one single maximum at  phase 0.5 when irradiation from the compact object plays a major role and dominates geometric effects (see e.g. \citealt{Davanzo}).
A single-maximum light curve is often observed in transient LMXBs in quiescence, which testifies to the strength of irradiation effects. Since 
the observed quiescent X-ray luminosity is not sufficient to account for the observed optical modulation, the energy released by a rotating magnetic dipole has been suggested as the main responsible factor for irradiation in quiescent LMXBs and AMXPs (\citealt{Burderi}; \citealt{Campana2004}).

The system IGR J00291+5934 is an AMXP that contains the fastest X--ray pulsar (1.67 ms) as compact object and has an orbital period of 2.46 hr \citep{Galloway2005}.
The system was first detected during an outburst on 2004 December 2 in an \textit{INTEGRAL} observation (\citealt{Eckert2004}; \citealt{Shaw2005}).
The companion star is thought to be a hot brown dwarf, derived for an isotropic distribution of inclinations and a NS mass of $2M_{\odot}$ \citep{Galloway2005}. 
The source distance has not been firmly determined to date (a lower limit of $\sim 4$ kpc has been given by \citealt{Galloway2005} by equating the long-term mass accretion rate deduced from the fluence observed in outburst to the expected gravitational-wave-driven mass transfer rate). 

IGR J00291+5934 was first observed during quiescence by \textit{Chandra} and \textit{ROSAT} in 2005 at a 0.5-10 keV luminosity level of $\sim 10^{32}$ erg s$^{-1}$ \citep{Jonker2005} and was later confirmed by XMM-Newton observations \citep{Campana2008}.

The optical counterpart was identified with an $ R\sim 17.4 \, \rm mag$ star during outburst with weak He II and H$ \alpha $ emission 
(\citealt{FoxKulkarni2004}; \citealt{Roelofs2004}; \citealt{Torres2008}; see also \citealt{Lewis2010}). 
The quiescent optical counterpart was discovered by \citet{Davanzo07} and \citet{Torres2008} as a faint $R=23.2\pm0.1$ object. Sinusoidal variability at the known orbital period with an $R$ and $I$ semi-amplitude of $\sim$ 0.2-0.3 mag suggested strong irradiation \citep{Davanzo07}. Based on $VRIJHK$ photometry, the required irradiating luminosity is $\sim 5 \times 10^{33}$ erg s$^{-1}$, which is much higher than the observed X-ray luminosity in quiescence \citep{Davanzo07}.
\citet{Jonker2008} observed IGR J00291+5934 in quiescence for several orbital periods in the $I$ band. They found evidence 
of strong flaring (up to 1 mag). After removal of the strongest optical flares, a weak orbital modulation is still present.

In this paper we investigate the quiescent optical emission of IGR J00291+5934 in more detail through observations performed with the 10.4m Gran Telescopio Canarias (GTC). We obtained $ g,r,i $ data over more than one orbital cycle (see Sect. 2). Strong flaring variability is also apparent in our data. We derive an orbital modulation and spectral energy distributions close to a flare peak and during an interval of 
low flaring activity (Sect. 3). The discussion is presented and conclusions are drawn in Sect. 4.

\section{Observations and data analysis}
  
%
The source IGR J00291+5934 was observed in quiescence during the night of 2014 August 31, using the Optical System for Imaging and low-Intermediate-Resolution Integrated Spectroscopy (OSIRIS; \citealt{Cepa2000}) camera mounted 
on the GTC with the $ g,r,i $ filters (PI: A. Papitto). The seeing was good and remained $ < 1'' $. Fifteen images in each $ g,r,i $ filters were taken, with an exposure time of 240 s, 
180 s, and 150 s, respectively (and cycling among filters), with the aim of covering the 2.48 hr orbital period.

Image reduction was carried out using the standard procedures, which consist of subtracting an average bias frame and 
dividing by a normalised flat frame. Aperture photometry was performed on the field using the {\tt daophot} task \citep{Stetson1987} 
for all the sources in the field. The photometric calibration was made against the Stetson standard field star PG1528 
\citep{Stetson2000}. We performed differential photometry with respect to a selection of isolated and non-saturated 
field stars, in order to minimise any possible systematic effect. 

\section{Results}\label{sec_results}

\begin{figure*}
\begin{center}
\includegraphics[scale=0.45]{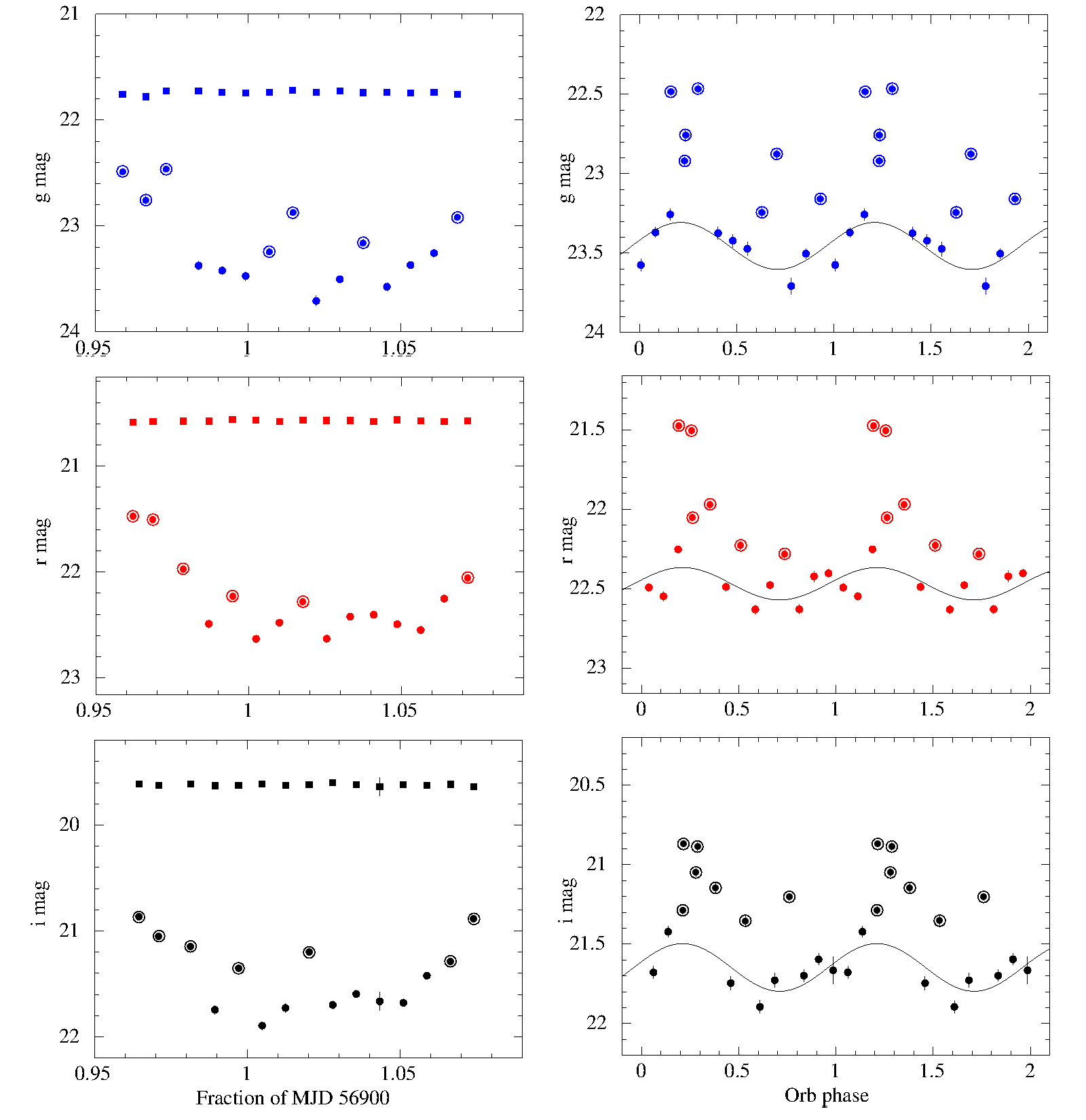}
\caption{\textit{Left panel}: From top to bottom, $g$, $r$, $i$ light curves (mag vs. MJD) of the system IGR J00291+5934 (dots) and of a calibration star (squares). Circled dots represent the possible flaring activity. \textit{Right panel}: From top to bottom, $g$, $r$, $i$ light curves (mag vs. orbital phase) of the system IGR J00291+5934. Circled dots represent the points corresponding to the flaring activity. Superimposed on the light curves we sketch the constant plus sinusoidal fit of the data, carried out after filtering the flares and imposing the same phase on all the light curves and with the period fixed to 1. In the $r $ band, the point corresponding to phase 0.73 is considered as a  flare for continuity with the other bands. Phase 0 corresponds to the inferior conjunction of the companion star. We evaluated the phases based on the X-ray ephemerides of \citet{Galloway2005}. Two periods of the system are drawn for clarity.}
\label{lc}
\end{center}
\end{figure*}

As can be observed in Fig. \ref{lc}, the light curves of IGR J00291+5934 are plagued by strong flaring activity that is prominent in all the observed bands with a small preference for shorter wavelengths,
and with up to 1 mag variability. We tried to filter the flares by determining a magnitude threshold by
visual inspection for each band. On this we define the flaring activity (i.e. $g=23.25$, $r=22.25,$ and $i=21.40$). Unfortunately, a detailed analysis such as reported in \cite{Jonker2008} is not possible with our dataset because of the poor temporal resolution. After this rough filtering, we tried to recover the possible orbital modulation, hidden by the flares (see Fig. \ref{lc}). We fit the three folded light curves together, imposing them to have the same orbital phase; the reduced $\chi^2$ of the best fit is not compelling ( $\chi^2/dof\sim 187.2/18$), however,
probably due to a combined effect of the very small errors and residual flaring activity. The results of the fits are reported in Table \ref{phot}. 

The source IGR J00291+5934 appears brighter than in \citealt{Davanzo07} and \citealt{Jonker2005} (0.6-0.8 mag and 0.2-0.3 mag, respectively). This effect is confirmed even when taking the different filters used by these authors into account (see Sec. \ref{disc_sec}). In addition, the phase of the maximum is not at 0.5 (superior conjunction of the companion star) but at $0.21\pm0.02$.
A small shift was also found by \citet{Jonker2005}, who found the maximum at phase $0.34\pm0.03$. 

The semi-amplitude of the modulations (Table \ref{phot}) is in line with previous results by \citet{Davanzo07}, who found a modulation of $0.22\pm0.09$
by combining $R$ and $I$ data, but not with \citet{Jonker2005}, who found a very small amplitude in the $I$ band of $0.06\pm0.01$.
However, we emphasise that all these results are strongly dependent on the flare filtering.

\begin{table*}
\caption{Results of the photometry of IGR J00291+5934. The mean magnitudes are not corrected for reddening. The reddening is obtained starting from the $ N_{\rm H} $ value reported in \citet{Davanzo07} (last column). 
Errors are indicated at the $90 \%$ c.l. }             
\label{phot}      
\centering         
\begin{tabular}{c c c c c c}  
\hline\hline                
Filter  & $\lambda_{\rm c}$ & Total Exposure& Mean         & Amplitude & $A_{\lambda}$\\     
           & (\AA)                           &   (s)                    & Magnitude &                   &    \\ 
\hline                       
   $g$ &$4770 $ & $3600$ & $ 23.45 \pm 0.02$ & $0.15\pm0.04$  &  $2.73 \pm 0.07$ \\
   $r$ &$ 6231$ & $2700 $ & $ 22.47 \pm 0.01 $ &$0.10\pm0.02$  & $1.88 \pm 0.07$ \\
   $i$ &$ 7625$ & $ 2250$ & $ 21.65 \pm 0.02$ &  $0.15\pm0.03$  &$1.38 \pm 0.07$ \\
\hline                                
\end{tabular}
\end{table*}

\begin{figure}
\begin{center}
\includegraphics[scale=0.3]{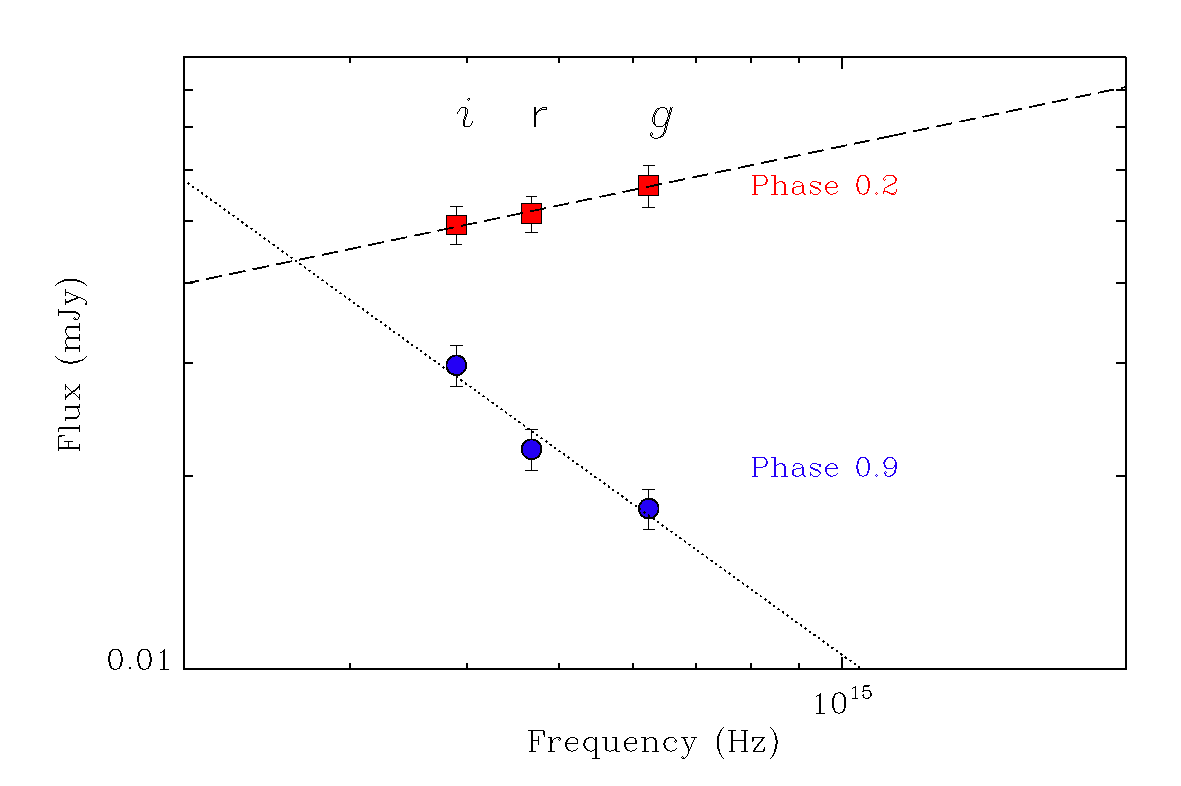}
\caption{Spectral energy distributions (SED) of the system IGR J00291+5934 in two different epochs: phase 0.2 
(red sqares) and phase 0.9 (blue dots). Phase 0.2 corresponds to a strong flare activity for the system. 
The fits of the two SEDs with a power law are shown (dashed and dotted lines, phase 0.2 and 0.9, respectively). Opposite behaviours can be observed ($ \alpha=0.31 \pm 0.35 $ 
and $ \alpha=-1.06 \pm 0.35 $, respectively). All the points are corrected for reddening, the parameters of which are reported in Table \ref{phot}.}
\label{SED}
\end{center}
\end{figure}

In order to investigate the nature of the flaring activity, we extracted the spectral energy distribution (SED, $gri$) around phase 0.2 
(where a prominent flare is present) and around phase 0.9 (when the system is likely quiescent).
Data in the $r$ band were taken as reference, and $gi$ data (mag vs MJD) were linearly interpolated to estimate the flux at exactly the same phase
(Fig. \ref{SED}). Data were then corrected for intrinsic absorption evaluated in \citet{Davanzo07} and quoted in the last column of Table \ref{phot}. The two SEDs are markedly different.  
The flare SED can be fitted with a power law with index $ \alpha=0.31 \pm 0.35$ (1$\sigma$ c.l.), indicating that the emission 
peak is likely at higher frequencies than the $g$ band. The index is fully consistent with what is predicted for a multi-temperature 
accretion disc with $T>30,000$ K.
The ``quiescent'' SED instead decreases with frequency, with a power-law best fit of $ \alpha=-1.06 \pm 0.35 $.  
This is difficult to interpret and could arise at the downturn of an accretion disc spectrum, immediately after the peak
(in this case, a disc temperature of $T\sim 10,000$ K is needed), or it could be the tail of the emission coming from the companion star (a brown dwarf; \citealt{Galloway2005}).


\section{Discussion}\label{disc_sec}
We report a study of the optical ($ gri $) light curves of the LMXB IGR J00291+5934 performed during quiescence, following two studies in the NIR--optical of the quiescent light curves of the same system (\citealt{Davanzo07}; \citealt{Jonker2008}) that reported different results. 

\citet{Davanzo07} provided optical and NIR photometry of the source during quiescence, with data acquired in 2005, less than one year after the end of an outburst in 2004 (\citealt{Eckert2004}; \citealt{Shaw2005}), and found sinusoidal variability modulated at the source 2.46 h orbital period, with a semi--amplitude of 0.2--0.3 mag, in both the $R$ and the $I$ band. The light curves peaked at phase $\sim 0.5$, suggesting a strongly irradiated companion star, as expected for AMXPs. This conclusion was also supported by the modelling of the NIR/optical spectral energy distribution of IGR J00291+5934, consistent with an irradiated star only, with no need for a residual accretion disc.

In light of these results, \citet{Jonker2008} reported puzzling
findings.
In 2006 the source, still in quiescence, showed a strong flickering activity in the $ I $ band, with flares up to $ \sim $ 1 mag and duration of $ \sim 10 $ min, that hid a sinusoidal modulation that is mosst likely due to the irradiated companion star. When the flares were subtracted, \citet{Jonker2008} was finally able
to observe the sinusoidal modulation of the $ I $ band light curve, which was peaked at phase 0.34. Moreover, an overall (non-significant) brightening of the source of $ 0.5\pm0.2 $ mag was observed with respect to what was found by \citet{Davanzo07}, consistently with a somewhat enhanced activity of the source in 2006 with respect to normal quiescence (2005).


Our observations were acquired in 2014, eight years after those reported in \citet{Jonker2008}. During this time, the system underwent a long-lasting outburst ($\sim 100$ days; \citealt{Lewis2010}) in 2008 and then returned to quiescence, where it remained until the epoch of our observations. Similarly to what is reported in \citet{Jonker2008}, our light curves do not show a clear sinusoidal trend (Fig. \ref{lc}). In particular, a strong flickering activity is detected in all bands, above all in $ g $, with flares up to $ \sim 1 $ mag and duration of the order of tens of minutes (i.e. comparable to those observed by \citealt{Jonker2008}). 
After excluding all flares (Fig. \ref{lc}), the light curves show a hint of a sinusoidal modulation at the known orbital period, particularly in the $ i $ band, where the irradiated companion star is expected to contribute the most. The fit of the light curves with a constant plus sinusoid model resulted in a $ i $--band mean magnitude of $21.65\pm 0.02$, which corresponds to $I=21.18\pm0.02$, according to the transformations reported in \citet{Jordi2006}. This shows that the system brightened further by $0.72\pm0.09$ mag in 2014 with respect to 2006 \citep{Jonker2008}, with a significance of $7.8\sigma$. Moreover, the orbital phase corresponding to the maximum brightness is $0.21 \pm 0.02 $ in all bands, significantly shifted with respect to what is expected for an irradiated object (phase 0.5; \citealt{Davanzo07}). A similar phase shift was also detected by \citet{Jonker2008}. 

From its discovery in 2004 up to now, IGR J00291+5934 underwent three different outbursts, in 2004 (\citealt{Eckert2004}; \citealt{Shaw2005}), 2008 \citep{Lewis2010}, and 2015 (\citealt{Cummings2015}; \citealt{Sanna2015}; \citealt{Russell2015_ATel7837}; \citealt{Sanna2016}; \citealt{Ferrigno2016}; \citealt{Patruno2016}). After each outburst, the accretion disc that surrounds the compact object in an LMXB is expected to empty out, leaving the companion star alone as the only (or main) responsible
source for the NIR--optical emission. This can easily explain why after quiescence begins, an observer usually sees a sinusoidal (or ellipsoidal) modulated light curve in the NIR/optical. However, during quiescence the replenishment of the accretion disc can start again, thus explaining why in a great number of X--ray binaries during quiescence a contribution in the optical that
is due to a residual accretion disc is indeed observed (see e.g. \citealt{Davanzo}). In the case of IGR J00291+5934, it is possible that after the 2004 outburst the contribution of the accretion disc, at that time almost empty, to the optical emission was minimal, thus leaving the companion star to dominate the overall emission and explaining the sinusoidal light curves reported in \citet{Davanzo07}. 
As shown in \citet{Davanzo07}, the companion star was not able to account for the observed optical modulation, thus ruling out a strong contribution from the accretion disc. Our suggestion was then that the companion star was strongly irradiated by an additional energy source, which we tentatively identified in the relativistic wind of a spinning neutron star.

After that, the accretion disc continued replenishing, getting ready for the new outburst in 2008. Exactly 699 days before the outburst began, \citet{Jonker2008} observed the system again in the optical in 2006, and the contribution of the accretion disc was indeed no longer negligible: the system was brighter and showed flares. The same might have occurred even some years later: our observations in 2014 were made only 327 days before the 2015 outburst took place, thus possibly explaining why a strong flaring activity was detected, together with the brightening of the source with respect to the level of normal quiescence  \citep{Davanzo07}. A similar behaviour was also stated in \citet{Wang2013} for the AMXP SAX J1808.4-3658, for which a brightening of the accretion disc emission 1.5 months before the onset of a bright outburst was observed.


The observations carried out by \citet{Jonker2008} and in this work instead show a source with a higher luminosity and strong flaring. An orbital modulation in the optical band is barely visible in the data, even if it is detected at a level similar to the modulation observed by \citet{Davanzo07}. This suggests that the companion star contribution remained similar, but that
the disc contribution grew considerably.
With the growth of the accretion disc, an increase in X-ray luminosity might also be expected. This was not observed, however, as confirmed
by pointed Swift observations on 2014, August 16 (ID:00031258002). In particular, no source is detected in a 20-pixel radius around the position of the target, with a $3\sigma$ upper limit on the count rate of $9\times10^{-3}$ $\rm counts \, \rm s^{-1}$, which translates into an upper limit of $6\times 10^{-13} \rm erg \, \rm cm^{-2} \, \rm s^{-1}$ on the unabsorbed X-ray flux (for $N_{\rm H}=6\times10^{21} \rm cm^{-2}$ and a power law with index 2.4; \citealt{Torres2008}).
The two statements (larger disc emission in the optical and lack of X--ray emission) can be reconciled assuming some inefficient accretion mechanism, such as the propeller mechanism (see e.g. \citealt{Campana2016}).

The observation of a strong flaring activity (Fig. \ref{lc}) shows that the companion star of the system is not the principal player in the optical emission of the system. The observation of a consistent phase shift of the optical light curves (after the subtraction of the flares) with respect to what is expected for an irradiated AMXP (phase 0.5) is further evidence. In particular, the phase shift suggests that at least part of the modulation that is sketched in Fig. \ref{lc} may arise in a somewhat different location, like an asymmetry in an accretion disc (hot spot or
superhump), with a periodicity that is different from the orbital period. 
Our light curves seems to suggest a non--negligible contribution of a newly formed accretion disc. The same is also supported by the spectral energy distribution built during a flaring event (Fig. \ref{SED}), which is fully consistent with a multi-temperature accretion disc with $T > 30, 000 \rm K$ (see Sect. \ref{sec_results}).
A similar conclusion was also reached by \cite{Jonker2008}, who proposed a superhump as the possible cause for both the observed variability and the phase shift of their light curve. This is doubtful, however: although some conditions for the formation of a superhump are fullfilled in the case of IGR J00291+5934 (e.g. the mass ratio between the companion star and the compact object is $ \lesssim $ 0.3; \citealt{Whitehurst1991}), others are not. In particular, it is known that superhumps tend to develop mainly during outbursts, that is, when the accretion disc extension is the largest, and this is not the case for the observations reported in \cite{Jonker2008} and in this work. However, in some rare cases superhumps have been detected even during quiescence \citep{Neilsen2008}, which leaves this hypothesis still open.

Mass transfer instabilities, magnetic reconnection events, and X-ray reprocessing in the accretion disc (\citealt{Hynes2002}; \citealt{Zurita2003}) might in principle also explain a strong flaring activity in LMXBs. No enhanced X-ray activity has been detected in IGR J00291+5934 near the epoch of our observations in 2014. For this reason, X-ray reprocessing can be safely ruled out. 
It has to be noted that X--ray flares are observed in some transitional millisecond pulsars (XSS J12270-4859 and PSR J1023+0038; \citealt{DeMartino2013}; \citealt{Patruno2014}; \citealt{Bogdanov2015}). However, in these cases, the flaring activity is only detected in the X-rays and has different timescales than in IGR J00291+5934, suggesting
that a different mechanism is at play.


Instabilities in the mass transfer rate from the donor star might induce variability at the hot spot, for example. This is the point where accreting matter impacts the accretion disc. If this is the case, the flaring activity should be higher when the visibility of the hot spot is at its maximum, that is, at phase 0.8 \citep{Zurita2003}. Since at phase 0.8 our system seems to be quiet (Fig. \ref{lc}), however, we can exclude this possibility as well.

Magnetic reconnection events in the disc could also trigger the flares (\citealt{Hynes2002}; \citealt{Zurita2003}). In particular, differential motions in the accretion disc plasma may produce magnetic fields, and when the reconnection between vertical field lines of opposite signs occurs, some energy is dissipated, leading to the emission of flares. In this scenario, the duration of flares must then be related to the timescale of shearing in the accretion disc, and should be more efficient in the outer regions of the disc, where the amount of shearing is higher (see \citealt{Zurita2003}). If we assume that the duration of flares is linked to their location in the accretion disc, flares with the longest duration should then be the most significant. Since to test this scenario we are strongly limited by the duration of our observations and statistics, we can leave this possibility open.

\section{Conclusions}
We presented the results of optical $gri$ photometry of the accreting millisecond X--ray pulsar IGR J00291+5934 during quiescence. Observations were carried out with the GTC equipped with OSIRIS on 2014, August 31. 

The system displays a strong flaring activity in all bands, above all in the $g$ band. This flaring activity is comparable to the activitiy previously observed by \citet{Jonker2008} in intensity ($\sim 1 \rm mag$) and duration. When the flares were subtracted, we observed an indication of a sinusoidal modulation at the system orbital period. As in the case of the observations reported in \citet{Jonker2008}, a phase shift of the light curves with respect to phase 0.5 is detected. All our optical light curves are consistent with an enhanced activity of the source, with a significant $\Delta I$ with respect to the observations during quiescence reported in \citet{Davanzo07} of $0.7\pm 0.1\, \rm mag$. Finally, the spectral energy distribution built during a flare (at phase $\sim 0.2$) was fitted by a power law with index $\alpha=0.31\pm 0.32$, which is consistent with what is predicted for a multi-colour black body of an accretion disc with $T>30,000 \,\rm K$.
All these results can be explained when we consider that the principal player in the quiescent emission of IGR J00291+5934 of our dataset is not the companion star, as expected for a quiescent LMXB, but the accretion disc. The disc, after it was emptied out during the 2008 outburst, started replenishing in preparation to the next outburst (which occurred in July 2015). In this way, both the observed brightening of the source and the phase shift of the light curves can be explained, since the main optical emitter in the system might be an asymmetry in the disc, like a hot spot or a superhump. The controversial results reported in \citet{Davanzo07} and \citet{Jonker2008} could also be accounted for in this scenario: in the first case, the system was observed after the end of an outburst, which probably left the accretion disc almost empty, thus explaining why the typical modulation due to the irradiated companion star alone was observed. In the latter, instead, the system was preparing itself for the 2008 outburst; thus the accretion disc was no longer empty and probably strongly contributed to the quiescent optical emission of the system (as in the case of the 2014 dataset). 

According to this picture, we thus conclude that the observed 2014 flaring activity might indicate an accretion disc during quiescence. Magnetic reconnection events in the disc might be a likely possibility. Further multi--wavelength optical observations during quiescence, possibly over longer timescales, could shed light on the true origin of the quiescent flares of IGR J00291+5934.

\begin{acknowledgements}
PDA and SC acknowledge the Italian Space Agency (ASI) for financial support through the ASI-INAF contract I/004/11/. AP acknowledges support via an EU Marie Sklodowska-Curie Individual Fellowship under contract No. 660657-239 TMSP-H2020-MSCA-IF-2014, as well as fruitful discussion with the international team on ``The disk-magnetosphere interaction around transitional millisecond pulsars'' at ISSI (International Space Science Institute), Bern.
TMD acknowledges support via a Ram\'{o}n y Cajal Fellowship (RYC-2015-18148) and by the Spanish Ministerio de Economia y competitividad under grant AYA2013-42627. NR and DFT acknowledge support by the Spanish Ministerio de Economia y competitividad (MINECO) under grant AYA2015-71042-P and the Generalitat de Catalunya via grant SGR2014-1458 as well as the Programme CERCA. NR is supported by an NWO Vidi Award.
\end{acknowledgements}


\end{document}